\documentstyle[12pt,aaspp4]{article}

\begin{document}

\title{A Test of Pre-Main Sequence Evolutionary Models Across the
Stellar/Substellar Boundary Based on Spectra of the Young Quadruple GG
Tau\footnote{Based partly on observations with the NASA/ESA Hubble
Space Telescope, obtained at the Space Telescope Science Institute,
which is operated by the Association of Universities for Research in
Astronomy, Inc. under NASA contract No. NAS5-26555.}}
\author{Russel J. White\altaffilmark{2}, Andrea M.
Ghez\altaffilmark{2,}\altaffilmark{3},
Iain N. Reid\altaffilmark{4}, Greg Schultz\altaffilmark{2}}

\altaffiltext{2}{UCLA Division of Astronomy and Astrophysics, Los
Angeles, CA 90095-1562}

\altaffiltext{3}{Packard Fellow}

\altaffiltext{4}{Palomar Observatory, California Institute of
Technology, Pasadena, CA 91125}

\begin{abstract}

We present spatially separated optical spectra of the components of
the young hierarchical quadruple GG Tau.  Spectra of GG Tau Aa and Ab
(separation 0\farcs 25 $\sim$ 35 AU) were obtained with the Faint
Object Spectrograph aboard the Hubble Space Telescope.  Spectra of GG
Tau Ba and Bb (separation 1\farcs 48 $\sim$ 207 AU) were
obtained with both the HIRES and the LRIS spectrographs on the W. M.
Keck telescopes.  The components of this mini-cluster, which span a
wide range in spectral type (K7 - M7), are used to test both
evolutionary models and the temperature scale for very young, low mass
stars under the assumption of coeval formation.  Of the evolutionary
models tested, those of Baraffe et al. (1998) yield the most
consistent ages when combined with a temperature scale intermediate
between that of dwarfs and giants.  
The version of the Baraffe et al. models computed with a mixing length
nearly twice the pressure scale height is of particular interest as
it predicts masses for GG Tau Aa and Ab that are in agreement with
their dynamical mass estimate.

Using this evolutionary model and a coeval (at 1.5 Myrs) 
temperature scale, we find that the coldest component of the GG Tau
system, GG Tau Bb, is substellar with a mass of $0.044 \pm 0.006$
M$_{\rm \odot}$.  This brown dwarf companion is especially intriguing
as it shows signatures of accretion, although this accretion is not
likely to alter its mass significantly.  GG Tau Bb is currently the
lowest mass, spectroscopically confirmed companion to a T Tauri star,
and is one of the coldest, lowest mass T Tauri objects in the
Taurus-Auriga star forming region.

\end{abstract}


\section{Introduction}

The masses and ages of T Tauri stars, a class of young ($\lesssim
10^7$yrs), low mass (M $\lesssim$ 2 M$_\odot$) stars, are primarily
inferred from the comparison of their observationally determined
stellar temperatures and luminosities to
the predictions of theoretical pre-main sequence (PMS)
evolutionary models (e.g., D'Antona \& Mazzitelli 1994, 1997; Swenson et
al. 1994; Burrows et al. 1997; Baraffe et al. 1998).  Unfortunately,
these mass and age estimates are currently very uncertain because of
the uncertainties in the input physics
of the evolutionary models.  There are at best only minimal
constraints on the choice of opacities, the convection prescription,
and the proper treatment of the atmosphere (e.g., grey vs. non-grey).
As a consequence, evolutionary models are computed with a
variety of assumptions and predict discrepant masses and ages
when compared to a specific luminosity and temperature.  For example,
a 1 Myr, 0.08 M$_{\odot}$ star according to the Baraffe et al. (1998)
model has the same luminosity and temperature as a 5 Myr, 0.12
M$_{\odot}$ star according to the D'Antona \& Mazzitelli (1997) model.
Variations by a factor as large as 10 in age and 2 in mass are common
between the models.

An additional problem in determining the masses and ages of T Tauri
stars is that it is unclear whether a temperature scale similar to
dwarfs or giants is more appropriate for these moderately
over-luminous, young stars.  While the dwarf and giant temperature
scales are nearly identical near spectral type M0, the giant scale is
more than 400$^{\rm o}$K hotter than the dwarf scale by spectral type
M7 (e.g., Leggett et al. 1996 vs. Perrin et al. 1998).  Thus even with
accurate spectral types it is difficult to place T Tauri stars on an
HR diagram for comparison with evolutionary models.  For mid- and
late-M T Tauri stars, these temperature scale differences lead to
additional mass and age discrepancies by factors of a few.

Much of the uncertainty in the evolutionary models and the T Tauri
temperature scale is a consequence of the limited observational 
constraints at the youngest ages and at solar or lower masses.
Open clusters at an age $\gtrsim 100$ Myrs such as the Pleiades have been
used to test the later stages of evolutionary models; cluster members
spanning a wide range in mass should lie along a single isochrone for an
accurate evolutionary model using the proper temperature scale (e.g.,
Luhman 1998; Stauffer, Hartmann \& Barrado y Navascues 1995; Stauffer
et al. 1997).  While this test has had some success in distinguishing
between evolutionary models at these older ages, it is not possible to
conduct this test with the youngest clusters or T associations such as
Orion or Taurus-Auriga.  The spread in ages for cluster members in
these star forming regions is comparable to the age of the cluster
(e.g., Hillenbrand 1997; Luhman \& Rieke 1998; Luhman et al. 1998b).

Young binary stars provide a variety of methods for assessing the
validity of the masses and ages inferred from evolutionary models.
Binaries exhibiting orbital motion can be used to estimate the sum of
the stellar masses (Ghez et al. 1995; Roddier et al. 1995; Thi\'ebaut
et al. 1995; Simon, Holfeltz \& Taff 1996), double-lined spectroscopic
binaries provide accurate mass ratios (Lee 1992; Figueiredo 1997;
Prato 1998), and the rare eclipsing binaries yield both masses and
radii (Popper 1987; Claret, Gim\'enez \& Mart\'\i n 1995; Mart\'\i n
and Rebolo 1993; Corporon et al. 1994, 1996; Casey
et al. 1998).  Unfortunately, these types of binaries have 
offered little constraint on the earliest stages of PMS evolution
as the orbital solutions are too preliminary and/or the the
stellar components are near the zero age main-sequence (see review by
Mathieu et al. 1998).

The relative ages of young multiple systems offer an additional
constraint on evolutionary models and the T Tauri temperature scale.  
Unlike the youngest open clusters, the components of a multiple
system, which can be thought of as a mini-cluster, are expected
to have a negligible age difference.  The correct evolutionary model and 
temperature scale should yield the same age for all components.  
Here we pursue this test using the relative ages of the PMS quadruple
GG Tau in the nearby star forming region Taurus-Auriga (D = 140 pc;
Kenyon, Dobrzycka \& Hartmann 1994).  GG Tau is a hierarchical
quadruple comprised of two binary
stars (Figure 1).  The close pair, GG Tau A, with components Aa \& Ab
(separation 0\farcs 25 $\sim$ 35 AU) is separated by 10\farcs 1
from a wider pair, GG Tau B, with components Ba \& Bb (= GG
Tau/c; separation 1\farcs 48 $\sim$ 207 AU).  The GG Tau system is a
particularly useful system for testing evolutionary models since there
also exists a dynamical mass estimate for GG Tau A from kinematic
studies of its circumbinary disk (Dutrey, Guilloteau \& Simon 1994;
Guilloteau, Dutrey \& Simon 1998).

This test of evolutionary models requires accurately determined
stellar properties for each component, which can only be
extracted from spatially resolved spectra and photometry.  We have
therefore obtained spatially separated optical spectra of GG Tau Aa
and Ab using the Hubble Space Telescope (section 2.1.1), and
of GG Tau Ba and Bb with LRIS and HIRES on the 10-m W. M. Keck
telescopes (sections 2.1.2 \& 2.1.3).  These spectra are used in
conjunction with high spatial resolution photometry to determine the
spectral type, luminosity and accretion activity of each component
(section 3).  Constraints on evolutionary models and the inferred
stellar properties of the GG Tau components are discussed in section 4
and our conclusions are summarized in section 5.

\section{Optical Spectroscopy and Data Analysis}

\subsection{Faint Object Spectrograph on the Hubble Space Telescope}

The Faint Object Spectrograph (FOS) aboard the Hubble Space
Telescope (HST) was used to obtain spatially resolved spectra of 
each component of the 0\farcs 25 binary GG Tau A on 1995 Nov 8.  
Each star was observed through the 0\farcs09 $\times$ 0\farcs09
aperture with the G570H (4600 - 6800 \AA) and the G400H (3250 - 4800
\AA) gratings, yielding a spectral resolution of R $\approx$ 1400 (cf.
Keyes et al. 1995).  Similar FOS observations of other close binary
systems indicate that scattered light contamination from a
nearby ($\sim 0\farcs3$) point source is negligible ($\lesssim 0.2$
percent), as is expected from the sharp COSTAR PSF (Keyes et al.
1995).  The HST spectra were initially calibrated by the FOS
calibration pipeline (cf. Keyes et al. 1997).  However, flat fielding
information for small aperatures was not initially available, thus the
spectra were later flat fielded using the appropriate sensitivity
functions provided by T. Keyes \& E. Smith (private communication),
which are now available in the calibration pipeline.

\subsection{HIRES at Keck}

High resolution spectra of GG Tau Ba and Bb were obtained on 1997 Dec
10 with the Keck I telescope using the High Resolution Echelle
Spectrometer (HIRES; Vogt et al. 1994).  With the 0\farcs86 slit
(HIRES Decker 'C1'), the instrument yielded 15 spectral orders from
6350 - 8730 \AA\, (with gaps between the orders) at a spectral
resolution of R $\approx$ 45,000.  The wavelength scale was determined
from a thorium-argon calibration lamp exposure.  Several mid-M
Hyades stars were also observed with this setup for spectral
comparison.

Spectra of both components of GG Tau B were obtained by aligning the
pair along the slit.  The spectra were binned by 2 in the spatial
direction, which yielded an effective plate scale of 0\farcs38/pixel.
The peaks of the spatial intensity distribution for each component
were clearly separated, although the wings overlapped.
In order to deconvolve the spectra accurately, a simple model
consisting of two Gaussians was fit to each 1D cut in the spatial
direction of the two dimensional spectra.  The full-width half maximum
of the best fit Gaussians, approximately equal to the seeing, was
0\farcs8.

\subsection{LRIS at Keck}

Low resolution spectra of the 1\farcs48 binary GG Tau B were obtained
1997 Dec 9 with the Keck II Low Resolution Imaging Spectrograph (LRIS;
Oke et al. 1995).  The observations
were conducted with a 400 l/mm grating and a one arcsecond slit.  This
setup yielded a spectral resolution of R $\approx$ 1300, spanning the
wavelength range from 6400 - 9400 \AA.  The wavelength scale was
determined from a Neon-Argon calibration lamp exposure.

A spectrum of the fainter component, GG Tau Bb, was obtained by
centering the slit on the Bb component with the slit aligned
perpendicular to the separation axis.  This slit
orientation was chosen to minimize the contamination from GG Tau Ba.
To estimate the resulting contribution from Ba, a second 'offset'
spectrum was obtained on the opposite side of Ba at the same
separation and slit orientation.  This offset spectrum, which is
dominated by scattered light from Ba, should not be contaminated by
the faint companion Bb on the opposite side of Ba.  We therefore
assign the offset spectrum to GG Tau Ba.

Due to uncertainties in positioning the slit, the intensity of
the offset spectrum was an overestimate of Ba's
contribution to Bb's spectrum; a direct subtraction (Bb spectrum minus
offset) resulted in a negative continuum at the blue end.  The general
appearance of Bb's spectrum and the offset spectrum were markedly
different (with Bb's spectrum characteristic of a cooler object) and
suggested only a minimal amount of contamination by Ba in Bb's
spectrum.  In order to quantify and to remove Ba's contribution,
the offset spectrum was scaled such that the subtracted
spectrum (Bb spectrum minus scaled offset) yielded TiO absorption
strengths at 7124 \AA, 7743 \AA\, and 8432 \AA\, identical to those
measured in the HIRES spectrum.  The subtraction removed only about 20
percent from Bb's continuum at 6500 \AA\, and 5 percent from Bb's
continuum at 9000 \AA.  Uncertainty in this subtraction led
to an increased uncertainty in the continuum level near 6500 \AA\, and
thus in the inferred equivalent width (EW) of H$\alpha$ at 6563 \AA\,
for Bb (section 3.2).  The inferred spectral type, however, is quite
robust (section 3.1).

\section{Results}

The primary goal of our spectroscopic analysis is accurate spectral
classification of each component, which includes determining their
spectral types, confirming that all components are indeed T Tauri stars
and establishing their T Tauri types (classical vs. weak).  Classical
T Tauri stars are thought to be accreting circumstellar material
and, consequently, exhibit strong Balmer series emission attributable 
to the high temperature regions generated from the accretion flow
(e.g., Basri \& Bertout 1989; Gullbring 1994).  Weak T Tauri stars are
presumed to be experiencing no accretion and the `weak' Balmer series
emission common in their spectra is attributed to enhanced
chromospheric
activity (e.g., Walter et al. 1988).  We use the strength of H$\alpha$
emission to distinguish classical from weak T Tauri stars, and a
classical/weak dividing value set by Mart\'\i n (1998);  T Tauri
stars are considered classical if their EW(H$\alpha$) $\ge$ 5 \AA\,
for K stars, $\ge$ 10 \AA\, for M0-M2 stars and $\ge$ 20 \AA\, for
cooler stars.  We characterize the excess optical
emission or veiling for each component, which is common in classical
T Tauri stars with large accretion rates; neglecting this excess
emission can affect both the inferred spectral type and the stellar
luminosity (e.g., Hartigan et al. 1991).  The radial and rotational
velocities of GG Tau Ba and Bb are also extracted from the HIRES
spectra.

\subsection{The Inferred Spectral Types}

The HST FOS spectra of GG Tau Aa and Ab are shown in Figure 2.
Their spectral types were established by comparison with spectral
standards from Montes et al. (1997) over the temperature sensitive
region 5700 - 6800 \AA.  This longer wavelength portion of
each spectrum was used for spectral classification since it suffers
the least from continuum excess emission common in classical T Tauri
stars (cf. Basri \& Batalha 1990; Hartigan et al. 1991; Gullbring et
al. 1998).  Additionally, this wavelength region is not strongly
sensitive to surface gravity, which can affect spectral classification
at even longer wavelengths for these moderately over-luminous stars.
Based on these comparisons, we classify Aa as
a K7 $\pm$ 1 and Ab as an M0.5 $\pm$ 0.5.  The
best fit dwarf spectra are also shown in Figure 2.  As can be seen in
the Figure, the CaH (6382 \& 6389 \AA; Kirkpatrick, Henry \& McCarthy
1991) feature is weaker in both Aa and Ab than in the comparison dwarf
and is indicative of a lower surface gravity.

The LRIS spectra of GG Tau Ba and Bb are shown in Figure 3.  Sections
of their HIRES spectra near Li I 6708 \AA, H$\alpha$ 6563 \AA\, and K
I 7699 \AA\, are shown in Figure 4.  The spectral types for Ba
and Bb were established by comparison of the LRIS spectra with
spectral standards from Kirkpatrick et al. (1991) and
Kirkpatrick, Henry \& Irwin (1997) over the temperature sensitive
region 6400 - 7600 \AA.  Based on these comparisons, we classify
Ba as an M5 $\pm$ 0.5 and Bb as an M7 $\pm$
0.5.  The best fit dwarf and giant spectra are shown in Figure 3, and
are roughly identical over this temperature sensitive region except
near the gravity sensitive CaH band at $\sim$ 7000 \AA\, (Kirkpatrick et
al. 1991).  Photospheric absorption features in the HIRES spectra are
consistent with these inferred spectral types.

As with GG Tau Aa and Ab, the surface gravities of GG Tau Ba and Bb
appear less than that of a dwarf.  This is particularly
striking in the LRIS spectrum of Bb.  As Figure 3 shows, the strength
of the gravity sensitive CaH band at $\sim$ 7000 \AA\, for Bb is
intermediate between that of giants and dwarfs, while other gravity
sensitive features (K I at 7665 \& 7699, Na I at 8183 \& 8195;
Kirkpatrick et al. 1991; Mart\'\i n, Rebolo \& Zapatero-Osorio 1996;
Schiavon et al. 1997) are weak and are much more giant-like than
dwarf-like.  Figure 4 shows the K I absorption at 7699 \AA\, in detail
for both Ba and Bb.  It is notably narrower in the T Tauri stars than
in the Hyades M5 dwarf RHy 83 (Reid 1993), again
indicative of a lower surface gravity.  These results are consistent
with the giant-like characteristics of other mid- to late-M T Tauri
stars (Basri \& Marcy 1995; Luhman, Liebert \& Rieke 1997; 
Brice\~no et al. 1998; Luhman et al. 1998a; Luhman \& Rieke 1998;
Neuh\"auser \& Comer\'on 1998) and strengthens our conclusion that
these are very young PMS objects.

\subsection{The T Tauri Type and Continuum Excesses}

All four GG Tau components exhibit prominent H$\alpha$ emission
(Figures 2, 3, 4; Table 1).  For the close pair, this emission is
much stronger in Aa than Ab (EW[H$\alpha$]$_{\rm Aa}$ = $-57
\pm 2$ \AA\, and EW[H$\alpha$]$_{\rm Ab}$ = $-16 \pm 1$ \AA).  For the
wider pair, the EW[H$\alpha$] measured from
both the LRIS and the HIRES spectra give similar values for GG Tau Ba
($-21 \pm 1$ \AA\, \& $-24 \pm 1$ \AA).  However, the value measured for Bb
varied by a factor of two ($-43 \pm 4$ \AA\, \& $-20 \pm 1$ \AA), which
occurred over a 1 day timescale.  Rapid variations such as this have
been previously observed for T Tauri stars (e.g., Johns-Krull \& Basri
1997), and may be quite common for late-M T Tauri stars where
relatively small changes in the H$\alpha$ emission can result in
significant changes in the observed EW because of the relatively low
continuum flux.  

Strong Li I absorption at 6708 \AA, a signature of extreme youth,
was detected in the HIRES spectra of both GG Tau Ba and Bb (Figure 4,
Table 1).  Because of the limited resolution of
the FOS spectra, no Li I was detected in the
spectra of GG Tau Aa or Ab (Table 1 gives upper limits).  However,
high resolution ground-based spectra of the unresolved pair, which are
known to be a physical pair based on orbital motion (Ghez et al. 1995;
Roddier et al. 1996), show strong Li I absorption
(EW[Li I] = 0.72 \AA; Basri, Mart\'\i n \& Bertout 1991).  
The low surface gravity features and Li I absorption, in conjunction
with the strong H$\alpha$ emission above the weak T Tauri star limit
(section 3) imply that all four components of the GG Tau system
are classical T Tauri stars.

The resolution of our FOS spectra is insufficient to determine any
continuum excess emission or veiling which may be
present in the spectra of Aa and Ab.  However, using higher resolution
ground-based spectra of the unresolved GG Tau A pair, Gullbring et al.
(1998) find GG Tau A exhibits only a small amount of excess emission.  
Specifically, they measure the ratio of excess flux to
photospheric flux, $r$, to be only 0.24 over the spectral region 4500 -
5100 \AA.  This excess emission, which can be 
attributed to the optically brighter primary ($\Delta$V = 2.9; section
2.2), is consistent with other continuum excess measurements for
GG Tau A (Basri \& Batalha 1990; Hartigan et al. 1991) implying that the
veiling is not highly variable and is usually low.  Since $r$ is
known to decrease towards longer wavelengths where the photosphere
becomes relatively brighter (Basri \& Batalha 1990; Hartigan et al.
1991), we expect little effect on the inferred spectral types; large
veilings at levels greater than $r \sim 0.4$ at 6300 \AA\, are
needed to alter the observed spectral type for Aa (K7) by one spectral
subclass (M0), which is nevertheless within our uncertainties.

Additionally, since Balmer series emission is known to be correlated
with continuum veiling (Basri \& Batalha 1990; Hartigan et al. 1990), 
the relatively weak Balmer series emission for Ab, which is only
modestly above the weak T Tauri star limit (section 3), suggests that
it experiences little or no optical veiling as well.

With high resolution spectra of both GG Tau Ba and Bb, continuum
excess emission can be measured directly from these spectra once their
radial velocities and rotational velocities are established.  The
radial velocities for Ba and Bb were determined via cross-correlation
with Gl 876, Gl 447 (Delfosse et al. 1998) and Hyades M dwarfs (Reid
1999).  We measure radial velocities of $16.8 \pm 0.7$ km/s for Ba and
$17.1 \pm 1.0$ km/s for Bb.  The rotational velocities ($v$sin$i$) for
Ba and Bb were determined from the width of the peak in their
cross-correlation with the slowly rotating ($<$ 2 km/s) mid-M dwarfs
Gl 876 and Gl 447 (Delfosse et al. 1998).  We find a $v$sin$i$ of
$9 \pm 1$ km/s for Ba and $8 \pm 1$ km/s for Bb.  The uncertainties in
both the radial and rotational velocities are determined from the
scatter in these estimates over several orders.

The spectra of GG Tau Ba and Bb were then compared to dwarf spectra
of similar spectral type and $v$sin$i$ to identify the presence of
continuum veiling (Hartigan et al. 1989).
The T Tauri spectra were modeled as a standard stellar spectrum plus a
constant level of excess emission.  Orders with no prominent
telluric absorption features were divided up into 20 \AA\, bins and the
optical excess emission within each bin was determined by minimizing
the $\chi^2$ of the model fit.  We found no detectable veiling for
either GG Tau Ba or Bb over nearly the entire wavelength coverage of
the spectra.  The veiling upper limits at 6500 \AA\, are $r < 0.1$ for
Ba and $r < 0.25$ for Bb, while at 8400 \AA\, the limits are $r < 0.05$
for Ba and $r < 0.1$ for Bb (where $r =$ F$_{\rm excess}$ / F$_{\rm
photosphere}$).  These veiling upper limits can be used to place upper
limits on the mass accretion rates.  If it is assumed that veiling
is caused by the gravitational energy released from accreting
material (Hartigan, Edwards \& Ghandour 1995; Gullbring et al. 1998),
then the mass accretion rate for both Ba and Bb must be less
than $10^{-9}$ M$_\odot$/yr.  Thus, although both GG Tau Ba and Bb do
show signatures of accretion (strong H$\alpha$ emission, NIR excess
emission), this accretion is not likely to alter their final masses
significantly.

\subsection{Extinction and Luminosity Estimates}

The spatially resolved photometric measurements for the four
components of GG Tau are shown in Figure 5 and listed in Table 2.  The
optical values are from the HST measurements of Ghez, White \&
Simon (1997), which have been transformed onto the more standard
Johnson-Cousins system (cf. Ghez et al. 1997).  The NIR measurements
are the median of spatially resolved JHKL photometry for GG Tau
compiled from both the literature and recent observations (White \&
Ghez 1999).

With spatially resolved photometry and spectral types, the
line-of-sight extinction and stellar luminosity of each component can
be determined from standard spectral type, color and bolometric
correction relations.  Much like the relative temperatures of
dwarf and giant stars, the colors of late-K and early-M dwarfs and
giants are similar, but by mid-M spectral types the optical
colors of giant stars begin to become bluer than the colors of dwarf
stars (cf. Bessell \& Brett 1988).  
Since dwarf colors are better established than giant colors, we adopt
dwarf colors in our analysis and give a simple argument for their
preference based on the derived extinctions below.
Specifically, we adopt the dwarf color relations of
Bessell \& Brett (1988) and  Bessell (1991) for the K7 star and that
of Kirkpatrick \& McCarthy (1994) for M0 and cooler stars.  We also
use the V-L color relations of Kenyon \& Hartmann (1995) for all
spectral types.

A mean line-of-sight extinction is determined for each component from
E(V-R$_{\rm c}$), E(R-I$_{\rm c}$) and E(I$_{\rm c}$-J), using the
extinction law of Rieke \& Lebofsky (1985) and color excess relations
of Hillenbrand (1997) and Luhman \& Rieke (1998).  We assign the
half-range of these three estimates as the uncertainty in the
extinction.  We note that if
giant colors (Th\'e, Steenman \& Alcaino 1984; Bessell \& Brett 1988)
are used to derive the extinction, the
half-range uncertainty estimates are considerably larger, especially
for the M5 and M7 components (A$_{\rm V}$ error for Ba[M5] = 0.37
magnitudes for dwarf colors vs. 0.69 magnitudes using giant colors and
A$_{\rm V}$ error for Bb[M7] = 0.24 magnitudes for dwarf colors vs.
2.12 magnitudes for giant colors).  This suggests that the intrinsic
photospheric colors of these T Tauri stars may indeed be more similar
to dwarfs than giants.

The luminosity is then derived from the reddening corrected I band
measurements and a bolometric correction, assuming a distance of 
140 pc, the distance to the Taurus-Auriga star forming region
(Elias 1978; Kenyon, Dobrzycka \& Hartmann 1994; Preibisch \& Smith
1997).  Typical uncertainties in the distance to Taurus-Auriga are 10
pc, although the large spatial extent of the star forming region on
the sky suggests that the uncertainty in the distance to any
particular star could be as large as 30 pc.  The effect of errors in
our assumed distance are discussed in section 4.2.

The bolometric corrections from Bessell (1991) are used for stars
hotter than spectral type M3 while the values from Monet et al. (1992)
are used for spectral type M3 and cooler.  While bolometric
corrections for early M stars are reasonably well established, their
remains considerable uncertainty in the bolometric corrections for
late-M stars (Monet et al. 1992; Kirkpatrick et al. 1993).  We assign
a generous uncertainty in our adopted bolometric corrections of 0.05
magnitudes for Aa and Ab, and 0.1 and 0.2 magnitudes for Ba and Bb,
respectively.  The uncertainty in the luminosity is determined from
the convolved uncertainty in the bolometric correction, the extinction
and the I band photometry.  These values are listed in Table 1.  It is
worth noting that the close pair, GG Tau Aa and Ab with a projected
separation of only 35 AU, differ in visual extinction by 2.5
magnitudes.  These extinction values are consistent with the reddened
photospheric continua observed in the FOS spectra.  The extinction
difference is most likely due to the differences in the local
distribution of circumstellar material for each component.  This
demonstrates the need for obtaining spatially separated spectra of all
components within a young multiple system in order to correctly
determine their stellar and circumstellar properties.

\section{Discussion}

\subsection{Evidence for Coeval Formation}

The four components of the GG Tau system appear to represent a
physical quadruple.  The GG Tau A and GG Tau B pairs have a projected
separation of
only 1400 AU.  This separation is considerably less than the size of a
typical cloud core ($\sim$ 20,000 AU) and therefore on a scale
consistent with that expected from core fragmentation models which are
thought to produce multiple systems (e.g., Bonnell \& Bastien 1993;
Boss 1996; Sigalotti \& Klapp 1997).  All components exhibit classical
T Tauri star signatures consistent with a similar stage of early
evolution.  More directly, GG Tau Aa and Ab are known to be a physical
pair based on orbital motion studies (Ghez et al. 1995; Roddier et al.
1996).  The HIRES radial velocity measurements of both GG Tau Ba and
Bb are in good agreement with the radial velocity measurements for GG
Tau A of $17.9 \pm 1.8$ by Hartmann \& Stauffer (1989), and thus are
consistent with a comoving system.  Millimeter maps of the ring around
GG Tau A show a well defined extension towards GG Tau B, which may be
interpreted as a tidal distortion due to GG Tau B (Koerner, Sargent \&
Beckwith 1993; Dutrey, Guilloteau \& Simon 1994).  It is also
important to consider that there are no other known T Tauri stars
within 30 arcminutes of the GG Tau system.  It thus
seems very unlikely that four closely spaced yet relatively isolated
classical T Tauri stars with similar radial velocities and stellar
properties would not be physically associated.

\subsection{A Test of PMS Evolutionary Models and the T Tauri
Temperature Scale}

The relative ages of the PMS quadruple GG Tau offer an
observational test of evolutionary models and the T Tauri temperature
scale.  The correct evolutionary model and temperature scale should
yield the same age for all components.  Currently, the GG Tau system
is uniquely suited for the relative age test.  Its multiple
components span a wide range in spectral type and have all been
resolved photometrically and spectroscopically.  Other systems with
apparently wide ranges in spectral type (e.g., UX Tau; Magazz\'u,
Mart\'\i n \& Rebolo 1991) typically have components that have not
been resolved spectroscopically (UX Tau B is a 0\farcs14 binary;
Duch\^ene 1998).

In order to place the components of GG Tau onto an H-R diagram for
comparison with evolutionary models, the observed spectral types need to
be converted into effective temperatures.  Unfortunately, the spectral
type - effective temperature relation for T Tauri stars is not well
known.  Since their surface gravity appears to be intermediate between
that of dwarfs and giants, the correct T Tauri temperature scale is
likely to be constrained between that of dwarfs and giants (Mart\'\i n
et al. 1994; Luhman et al. 1997).  We use these extremes as boundaries
for the range of plausible temperatures for T Tauri stars.

For the dwarf temperature scale, we use a K7 temperature of 4000$^{\rm
o}$K (Bessell 1991) and the M dwarf temperature scale of Leggett et al.
(1996) as fit by Luhman \& Rieke (1998), which is consistent with the
available observational constraints for dwarf stars (Luhman \&
Rieke 1998).  We use the giant temperature scale of Perrin et al.
(1998), which has been accurately established via interferometric
angular diameter measurements of cool giant stars.

In Figure 6, the components of the GG Tau system are plotted on
several H-R diagrams with both the dwarf (solid squares) and giant
(open
diamonds) temperature scales.  A dotted line connects the dwarf and
giant temperatures, identifying the range of plausible temperatures
for each component.  As the Figure illustrates, the dwarf and giant
temperature scales are essentially identical for spectral types near
M0, but diverge significantly for cooler spectral types.  The
errorbars plotted for each component correspond to the uncertainty in
the spectral type (section 3.1) and luminosity (section 3.3).

The locations of the GG Tau components define an empirical isochrone
which can be compared with PMS evolutionary models.  We conduct this
comparison using six popular PMS evolutionary models, which differ 
primarily in their prescription for convection, the assumed
opacities and the assumption of a grey or non-grey atmosphere.
The models we consider include (1) Swenson et al. (1994, model F,
hereafter S94), computed with a mixing-length theory for
convection, the opacities of Alexander (1992) and Iglesias,
Rogers \& Wilson (1992) and a grey atmosphere, (2) D'Antona
\& Mazzitelli (1994, hereafter DM94-MLT), 
computed with a mixing-length theory for convection, the
opacities of Alexander (1992) and Rogers \& Iglesias (1992)
and a grey atmosphere, (3) D'Antona
\& Mazzitelli (1994, hereafter DM94-CM), computed with
the Full Spectrum of Turbulence model for convection (Canuto \&
Mazzitelli 1992) and the same opacities and grey atmosphere
as DM94-MLT\footnote{D'Antona \& Mazzitelli (1994) also compute
evolutionary models using the Kurucz opacities.  We do not
consider these evolutionary models since the Kurucz opacities are
inadequate for temperatures $\lesssim 4000^{\rm o}$K where molecules
play an important role (D'Antona \& Mazzitelli 1997).}, (3) D'Antona
\& Mazzitelli (1997, hereafter DM97), also computed with the Full
Spectrum of Turbulence convection model and a grey atmosphere, but
with updated opacities (Alexander \& Ferguson 1994, Iglesias \& Rogers
1996), (5) Baraffe et al. (1998, hereafter B98-I),
computed with a mixing-length theory for convection with the
mixing-length equal to the pressure scale height in the atmosphere, 
updated opacities (Alexander \& Ferguson 1994, Iglesias \& Rogers
1996) and a non-grey atmosphere, and finally (6) Baraffe et al. (1998,
hereafter B98-II), which is identical to B98-I except that the
mixing length is increased by a factor of 1.9 for masses greater than
0.6 M$_\odot$\footnote{A factor of two change in the mixing length has
inconsequential effects on evolutionary models of mass $\lesssim$ 0.6
M$_\odot$ (Chabrier \& Baraffe 1997, Baraffe et al. 1998)}.  Each of
these models is computed with near solar abundances.

As shown in Figure 6, GG Tau Aa and Ab are coeval according to all
evolutionary models and either temperature scale, although as noted
above, the temperature scales are nearly identical for these spectral
types (K7 and M0.5).  We therefore use the isochrone defined by their
average age (dashed line in Figure 6) to test the evolutionary models
at cooler temperatures.

The S94 model yields the most discrepant ages (Figure 6a).  The
GG Tau A isochrone predicts temperatures considerably colder
than the range of plausible temperatures for GG Tau Ba and, with
modest extrapolation, GG Tau Bb.  The DM97 model exhibits a similar
problem (Figure 6b).  The coldest component, GG Tau Bb, is coeval
at a temperature consistent with the cool dwarf temperature, but its
slightly hotter companion, GG Tau Ba, is coeval at a temperature
well below even the dwarf temperature.
The DM94-MLT and DM94-CM models appear moderately more successful.
Both yield coeval ages (Figure 6c \& 6d), but require a very cool
dwarf temperature at M5 and then a considerable jump of more than a
200$^{\rm o}$K from the dwarf scale at M7.

The B98-I and B98-II models provide the most consistent ages
with a temperature scale intermediate between that of dwarfs and
giants (Figures 6e \& 6f).  For the B98-I model, a temperature scale
hotter than the dwarf scale by $\sim 160^{\rm o}$K is needed to make
both Ba and Bb coeval, resulting in effective temperatures of
3160$^{\rm o}$K and 2840$^{\rm o}$K, respectively.  The B98-II model
also suggests a temperature scale moderately hotter than the dwarf
temperature scale, with perhaps some spectral type dependence; the M5
is coeval when 40$^{\rm o}$K hotter than the dwarf temperature while
the M7 is coeval when 140$^{\rm o}$K hotter than the dwarf
temperature, resulting in effective temperatures of 3050$^{\rm o}$K
and 2820$^{\rm o}$K for Ba and Bb, respectively.

It is important to note that these results are not notably affected by
the uncertainties in the distance to GG Tau, even if significant
errors (30 pc) are assumed.  This is because evolutionary models are
primarily vertical in the H-R diagram, and the isochrones are roughly
parallel, for these masses at a young age.  Thus while changes in
distance can notably affect the absolute age, there is much less of an
effect on the relative age.  

The success or failure of these evolutionary models in yielding
consistent ages may identify inadequacies in the input physics of 
some evolutionary models.  The earliest models (S94, DM-MLT94 \&
DM-CM94) all assume a grey atmosphere and similar but slightly
outdated sets of opacities, but differ in their convection
prescription.  Their failure to yield coeval ages using a consistent
temperature scale suggests that a grey atmosphere and these older
sets of opacities are not appropriate for young stars, independent of
the convection prescription.  It is also worth noting that these
models poorly match the empirical Pleiades isochrone for masses below
$\sim 0.5$ M$_\odot$, although newer evolutionary models are
moderately more successful (Stauffer et al. 1995; Luhman 1998;
D'Antona \& Mazzitelli 1997).  However, the DM97 evolutionary model
computed with updated opacities does not yield a coeval age,
which suggests difficulties in the model other than the choice of
opacities.  Given the success of both the B98-I and B98-II models,
it appears that the assumption of a non-grey atmosphere is
an important condition for success.  This is not a surprising result.
Molecules become stable in the atmosphere at these low effective
temperatures ($\lesssim 5000^{\rm o}$K), and constitute a significant
source of absorption.  Since molecular absorption coefficients are
strongly depend upon wavelength, a grey atmosphere approximation is
not expected to be valid (Chabrier \& Baraffe 1997).

Although the relative age test cannot distinguish between the two
Baraffe et al. (1998) models, a dynamical mass estimate for GG Tau A
exists and offers an additional constraint.  Orbital velocity 
measurements of the circumbinary disk surrounding GG Tau A imply a
combined stellar mass (Aa plus Ab) of $1.28 \pm 0.08$ M$_\odot$
(Dutrey et al. 1994; Guilloteau et al. 1998).  For comparison, the
B98-I model implies a total mass for GG Tau A of $2.00 \pm 0.17$
M$_\odot$, while the B98-II model implies a total mass of $1.46
\pm 0.10$ M$_\odot$ (section 4.3).  The dynamical mass measurements
are clearly in much closer agreement with the B98-II models.  For
completeness we also report that the sum of the masses inferred for Aa
and Ab from S94 is $1.08 \pm 0.22$ M$_\odot$, from DM94-MLT is $1.12
\pm 0.16$ M$_\odot$, from DM94-CM is $0.81 \pm 0.17$ M$_\odot$ and
from DM97 is $0.80 \pm 0.14$ M$_\odot$.  The successful B98-II model
is also reasonably consistent with the available mass constraints for
GM Aur and DM Tau based on dynamical studies of their circumstellar
disks (see Appendix).  This agreement suggests that convective mixing
lengths nearly twice the pressure scale height are more appropriate
for young stars than mixing lengths equal to the pressure scale height.  
Alternative prescriptions for convection such as the Full Spectrum of
Turbulence method (Canuto \& Mazzitelli 1992) will hopefully also be
considered in future evolutionary models which incorporate a non-grey
atmosphere.  It is worth noting though that the two models computed
with this more complicated convection prescription (DM-CM94 \& DM97)
consistently predict masses which are systematically less than the
dynamical mass estimates for GG Tau A (given above), GM Aur and DM Tau
(Appendix).  We suggest that future evolutionary codes consider
convective scales which result in masses consistent with dynamically
inferred masses.

\subsection{The Inferred Stellar Properties:  A Substellar Companion}

Since the B98-II evolutionary model is the most consistent with the
available observational constraints, we use this model and the implied
coeval temperature scale, at an age of 1.5 Myrs, to infer the
stellar masses (Table 1).  The uncertainties in mass for Aa and Ab are
determined from the uncertainty in the spectral type, assuming a dwarf
temperature scale.  
The uncertainties in the mass of Ba and Bb are determined from the
uncertainty in their luminosity and the assumption that they are
constrained to lie on the average age isochrone of Aa and Ab.  Of
particular interest is the lowest mass component, GG Tau Bb, with a
mass of 0.044 $\pm$ 0.006 M$_\odot$.  This mass is well below the
hydrogen burning minimum mass of $\sim 0.075$ M$_\odot$ (Chabrier \&
Baraffe 1997), and can therefore be considered a young brown
dwarf.  As can be seen by comparing the B98-I and B98-II models,
changes in the assumed isochrone have little effect on the inferred
mass for Bb.  This is a consequence of the roughly steady state
burning of deuterium, which keeps $\sim 0.05$ M$_\odot$ objects at a
similar luminosity and temperature for nearly 5 Myrs (Baraffe et al.
1998).

With a spectral type of M7, GG Tau Bb is currently the coldest, lowest
mass, spectroscopically confirmed companion to a T Tauri star.  The
only confirmed T Tauri companion of similar spectral type is UX Tau C
(M5 - M6; Magazz\'u et al. 1991; Basri \& Marcy 1995).
Other low mass companions to T Tauri stars have been suggested
based on proximity and photometry (Tamura et al. 1998, Lowrance et al.
1999; Webb et al. 1999), but confirmation of their companionship and
mass requires spatially
resolved spectroscopy.  GG Tau Bb is also one of the coldest,
lowest mass T Tauri objects in the Taurus-Auriga star forming region.
Only a few objects of comparable mass (Brice\~no et al. 1998; Luhman
et al. 1998a) or possibly lower mass (Reid \& Hawley 1999) are known in
this region.
Moreover, only a handful of T Tauri objects with temperatures cooler
than GG Tau Bb are known in any star forming region (Hillenbrand 1997;
Luhman et al. 1997; Luhman et al. 1998b; Neuh\"auser \&
Comer\'on 1998; Wilking, Greene \& Meyer 1999; Reid \& Hawley 1999).


All components of the GG Tau system appear to support actively
accreting circumstellar disks.  These disks are inferred from the
strong Balmer series emission (Figures 2, 3 and 4) 
as well as UV and NIR excess emission (Figure 5)
characteristic of circumstellar accretion.  While circumstellar
accretion disks are common in binary systems of a few hundred AU
separation (Brandner \& Zinnecker 1997; Ghez et al. 1997; Prato 1998),
it is intriguing that the substellar companion also supports a
circumstellar disk.  It appears that young brown dwarf
companions of mass $\sim$ 0.05 M$_\odot$ can support a circumstellar
disk.  The similarities in the circumstellar properties of GG Tau Bb
compared to other low mass T Tauri binaries may also suggest, albeit
very speculatively, that brown dwarf companions of mass $\sim$ 0.05
M$_\odot$ form in the same fashion that it is believed young binary
stars form, from the fragmentation of a collapsing cloud core (e.g.,
Boss 1988).

\section{Summary}

The spatially resolved optical spectra of the young quadruple GG Tau
demonstrate that all components are classical T Tauri stars with
surface gravities intermediate between that of dwarf and
giants.  The similarities in their position (and relative
isolation within the cloud), kinematics, stellar and accretion
properties further establish these stars as a physical quadruple.

The components of this 'mini-cluster', which span from K7 to M7 in
spectral type, provide a test of evolutionary models and the
temperature scale for very young, low mass stars under the assumption
of coeval formation.  Of the evolutionary models tested, the Baraffe
et al. (1998) models, which are unique in their assumption of a
non-grey atmosphere, yield the most consistent ages using a
temperature scale intermediate between that of dwarfs and giants.  
The Baraffe et al. (1998) model computed with a mixing length nearly
twice that of the pressure scale height (B98-II) is the most
consistent with the dynamical mass estimates for GG Tau A, and
is also reasonably consistent with the dynamical mass
estimates for GM Aur and DM Tau.
This model suggests an age for the system of 1.5 Myrs and a
T Tauri temperature scale which diverges modestly from the dwarf
temperature scale:  $\sim$ M0 T Tauri
stars are consistent with the dwarf scale, M5 and M7 T Tauri stars are
roughly 40$^{\rm o}$K and 140$^{\rm o}$K hotter than the dwarf scale.
It is clear that PMS multiple systems with a wide range in stellar
masses similar to GG Tau can be used as a powerful test of
evolutionary models at this very interesting, but uncertain early
stage of stellar evolution.

Using the successful Baraffe et al. model (B98-II) with the
implied coeval temperature scale, we find that the GG Tau Bb component
is substellar with a mass of $0.044 \pm 0.006$ M$_{\rm \odot}$.  
This substellar companion is particularly intriguing as its large
H$\alpha$ emission and NIR excess emission suggest that it supports a
circumstellar accretion disk.  GG Tau Bb is currently the lowest mass,
spectroscopically confirmed companion to a T Tauri star, and is one of
the coldest, lowest mass T Tauri objects in the Taurus-Auriga star
forming region.

\acknowledgments

Some of the data presented herein were obtained at the W.M. Keck
Observatory, which is operated as a scientific partnership among the
California Institute of Technology, the University of California and
the National Aeronautics and Space Administration.  The Observatory
was made possible by the generous financial support of the W.M. Keck
Foundation.  Support for this work was provided by NASA through grant
NAGW-4770 under the Origins of Solar Systems Program and grant 
number G0-06014.01-94A from the Space Telescope Science Institute,
which is operated by AURA, Inc., under NASA contract NAS5-26555.  The
authors are grateful to W. Brandner, K. Luhman, J. Patience, D.
Popper, L. Prato,
R. Webb \& B. Zuckerman for helpful comments and discussions and to G.
Basri, D. Kirkpatrick \& K. Luhman for spectral standards.  We thank
D. Kirkpatrick and J. Stauffer for helping with
the LRIS observations and we greatly appreciate the assistance 
provided by the FOS instrument scientists T. Keyes and E. Smith.

\appendix

\section{Mass Constraints on Evolutionary Models from GM Aur and
DM Tau}

The single T Tauri stars GM Aur and DM Tau have mass estimates
from dynamical studies of their circumstellar material (Dutrey et al.
1998; Guilloteau \& Dutrey 1994, 1998).  Here we compare these mass
estimates to the predictions of evolutionary models.  
In order to place these two stars on the H-R diagram, we adopt
the spectral types of K7 for GM Aur (Gullbring et al. 1998, Basri \&
Batalha 1990, Hartigan et al. 1995) and M1 for DM Tau
(Hartmann \& Kenyon 1990).  These spectral types are converted to
effective temperatures using a dwarf temperature scale, which is
similar to the giant temperature scale for these
spectral types (section 4.2).  The line-of-sight extinctions and
luminosities are derived from the broad-band photometry reported in
Kenyon \& Hartmann (1995), assuming dwarf colors, a distance of 140 pc
and the methodology outlined in section 3.3.  For GM Aur (log\,T$_{\rm
eff}$ = 3.602, A$_{\rm V}$ = 0.10 mag, log(L/L$_\odot$) = $-0.25$),
the evolutionary models considered in section 4.2 predict masses of
0.70 M$_\odot$ (S94), 0.70 M$_\odot$ (DM94-MLT), 0.53 M$_\odot$
(DM94-CM), 0.51 M$_\odot$ (DM97), 
1.06 M$_\odot$ (B98-I) and 0.78 M$_\odot$ (B98-II).  The B98-II model
offers the best agreement with the dynamical mass estimates of $0.84
\pm 0.05$ M$_\odot$ (Dutrey et al. 1998).  For DM Tau (log\,T$_{\rm
eff}$ = 3.566, A$_{\rm V}$ = 0.63 mag, log(L/L$_\odot$) = -0.60), the
evolutionary models predict masses of 0.58 M$_\odot$ (S94), 0.48
M$_\odot$ (DM94-MLT), 0.43 M$_\odot$ (DM94-CM), 0.44 M$_\odot$ (DM97),
0.67 M$_\odot$ (B98-I) and 0.64 M$_\odot$ (B98-II).  Here the B98-II
model is less consistent the dynamical mass of $0.47 \pm 0.06$
M$_\odot$ (Guilloteau \& Dutrey 1998), while the various models by
D'Antona \& Mazzitelli (DM97, DM94-CM, DM94-MLT) are in agreement with
the dynamical mass.
However, DM Tau is known to be considerably veiled (Hartmann \& Kenyon
1990), which typically causes the optical spectrum to look like a
hotter star.  If the intrinsic spectral type for DM Tau is cooler by
one spectral subclass (M2), then the Baraffe et al. models (both B98-I
and B98-II are identical at this mass) predict a mass of 0.50 M$_\odot$
which is consistent with the dynamical mass.  We caution however that
GM Aur and DM Tau currently only offer weak mass constraints due to
the uncertainties in their distance and their spectral type.
Nevertheless, the B98-II model, in addition to being consistent with
the relative ages of the components of the GG Tau system, is also
reasonably consistent with the available dynamical mass constraints.

\newpage

\figcaption{GG Tau is comprised of four components whose relative
positions are shown in this 0.81$\mu$m HST/WFPC2 image.  North is up
and East is left.  This
quadruple system consists of two pairs of stars separated by 10\farcs1
(1400 AU).  The closest pair, GG Tau A, is a 
0\farcs25 (35 AU) binary and the wider pair, GG Tau B, is a 1\farcs48
(207 AU) binary.}

\figcaption{Hubble Space Telescope / Faint Object Spectrograph spectra
of GG Tau Aa and Ab normalized at 6500 \AA.  The Balmer series
emission is typical of classical T Tauri stars.  Dwarf spectra of
the same spectral type are also shown, displaced vertically, over the
wavelength region used for spectral classification.}

\figcaption{W. M. Keck II LRIS spectra of GG Tau Ba and Bb, normalized
at 7500\AA.  Both components show strong H$\alpha$ emission.  The
strength of some gravity sensitive features appears intermediate
between that of dwarfs and giants (e.g., CaH near 7000 \AA), although
the overall shape is best matched by a giant spectrum.}

\figcaption{Sections of the W. M. Keck I HIRES spectra of GG Tau Ba
and Bb near H$\alpha$, Li I, and K I.  The M5 Hyades dwarf RHy 83
(Reid 1993) is also shown
for comparison.  The spectra have been offset vertically for clarity,
but have the same relative scaling.  Both Ba and Bb show strong
H$\alpha$ emission and Li I absorption.  The wings of the K I feature
are much narrower for the T Tauri stars than for the dwarf which
is characteristic of a lower surface gravity.}

\figcaption{Spectral energy distributions of the GG Tau components
derived from HST optical observation and NIR speckle observations
(section 2.2).  The triangles represent upper limits.  A best
fit, reddened dwarf photosphere is also shown for each component.
The stars show NIR and UV excess emission commonly attributed to a
circumstellar accretion disk.}

\figcaption{The locations of the GG Tau components are shown on an HR
diagram using a dwarf temperatures scale (solid squares; Leggett et
al. 1996) and a giant temperature scale (open diamonds; Perrin et al.
1998).  Also plotted are current evolutionary models (see text).  The
four components of GG Tau define an isochrone with which the T Tauri
temperature scale and evolutionary models can be tested.  The Baraffe
et al. (1998) models provide the most consistent ages using a
temperature scale intermediate between that of dwarfs and giants
(starred points).}

\newpage
\begin{deluxetable}{llcccccc}
\tablecaption{Stellar Properties}
\tablewidth{0pt}
\tablehead{ \colhead{}
& \colhead{Spec.}
& \multicolumn{3}{c}{Equivalent Widths (\AA)}
& \colhead{A$_{\rm V}$}
& \colhead{log(L)} 
& \multicolumn{1}{c}{B98-II, 1.5 Myr} \nl
\colhead{GG Tau} 
& \colhead{Type} 
& \colhead{H$\alpha$} 
& \colhead{H$\beta$} 
& \colhead{Li I}
& \colhead{(mag)} 
& \colhead{(${\rm L_\odot}$)} 
& \colhead{Mass (${\rm M_\odot}$)} 
}
\startdata
Aa	& K7 & -57	& -29  & $< 1.0$ & $0.72 \pm 0.26$ &
$-0.074 \pm 0.067$ & $0.78 \pm 0.10$ \nl
Ab	& M0.5 & -16	& -7.0 & $< 1.3$ & $3.20 \pm 0.18$ &
$-0.146 \pm 0.051$ & $0.68 \pm 0.03$ \nl
Ba	& M5 & -20, -21	& \nodata & 0.56 & $0.55 \pm 0.39$ &
$-1.122 \pm 0.104$ & $0.12 \pm 0.02$ \nl
Bb	& M7 & -43, -20	& \nodata & 0.60 & $0.00 \pm 0.24$ &
$-1.822 \pm 0.104$ & $0.044 \pm 0.006$ \nl
\enddata
\tablecomments{The EW[H$\alpha$] measurements for Ba and Bb
are from LRIS and HIRES, respectively.}
\end{deluxetable}

\begin{deluxetable}{lllll}
\tablecaption{Optical and NIR Photometry}
\tablewidth{0pt}
\tablehead{ \colhead{Filter} 
            & \colhead{Aa}
            & \colhead{Ab}
            & \colhead{Ba}
            & \colhead{Bb} }
\startdata
U &		$13.72 \pm 0.04$ & $17.95 \pm 0.05$ & $19.11 \pm 0.12$
& $> 21.4$ \nl
B &		$13.62 \pm 0.02$ & $16.97 \pm 0.08$ & $18.36 \pm 0.09$
& $> 20.7$ \nl
V &		$12.30 \pm 0.02$ & $15.22 \pm 0.09$ & $16.97 \pm 0.07$
& $> 19.4$ \nl
R$_{\rm c}$ &	$11.32 \pm 0.02$ & $13.74 \pm 0.07$ & $15.36 \pm 0.06$
& $18.01 \pm 0.18$ \nl
I$_{\rm c}$ &	$10.46 \pm 0.02$ & $12.19 \pm 0.04$ & $13.39 \pm 0.02$
& $15.55 \pm 0.07$ \nl
J &		$9.24 \pm 0.21$ & $10.12 \pm 0.21$ & $11.48 \pm 0.04$
& $13.16 \pm 0.05$ \nl
H &		$8.27 \pm 0.25$ & $ 9.07 \pm 0.32$ & $10.63 \pm 0.09$
& $12.38 \pm 0.10$ \nl
K &		$7.73 \pm 0.16$ & $ 8.53 \pm 0.24$ & $10.20 \pm 0.13$
& $12.01 \pm 0.13$ \nl
L &		$6.83 \pm 0.04$ & $ 7.69 \pm 0.04$ & $ 9.32 \pm 0.03$
& $11.16 \pm 0.14$ \nl
\enddata
\tablecomments{Errors for UBVR$_{\rm c}$I$_{\rm c}$ and L 
represent measurement uncertainties; the components have
only been resolved once at these wavelengths.  Errors for the JHK
measurements represent the standard deviation of all spatially
resolved photometry (White \& Ghez 1998).  
Due to a red leak in the HST F336W filter, the transformed U-band
measurements given here are contaminated by roughly 10 percent
(Holtzmann et al. 1995; Ghez et al. 1997)} 
\end{deluxetable}

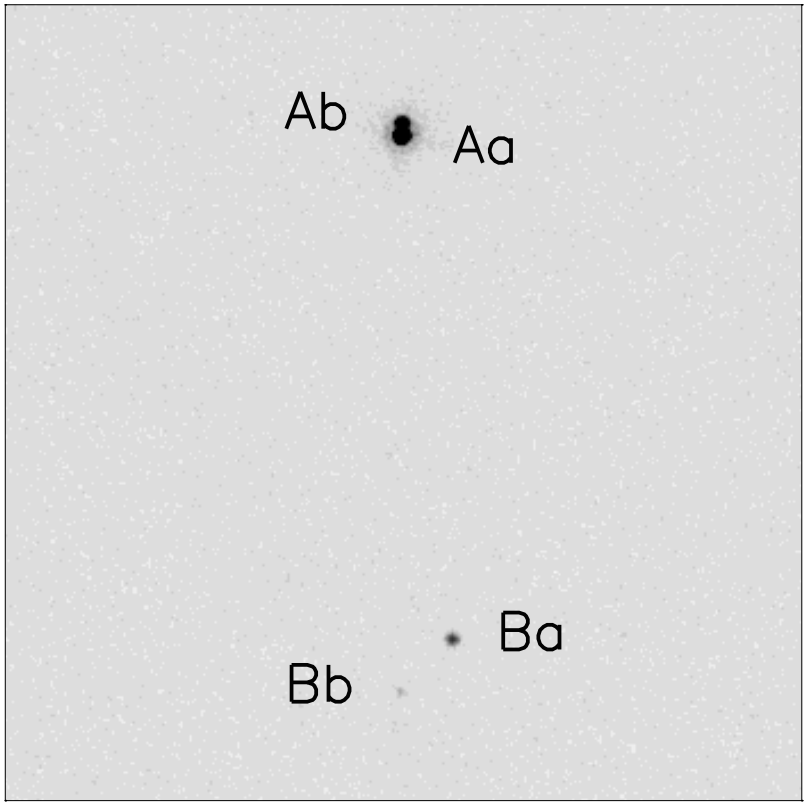
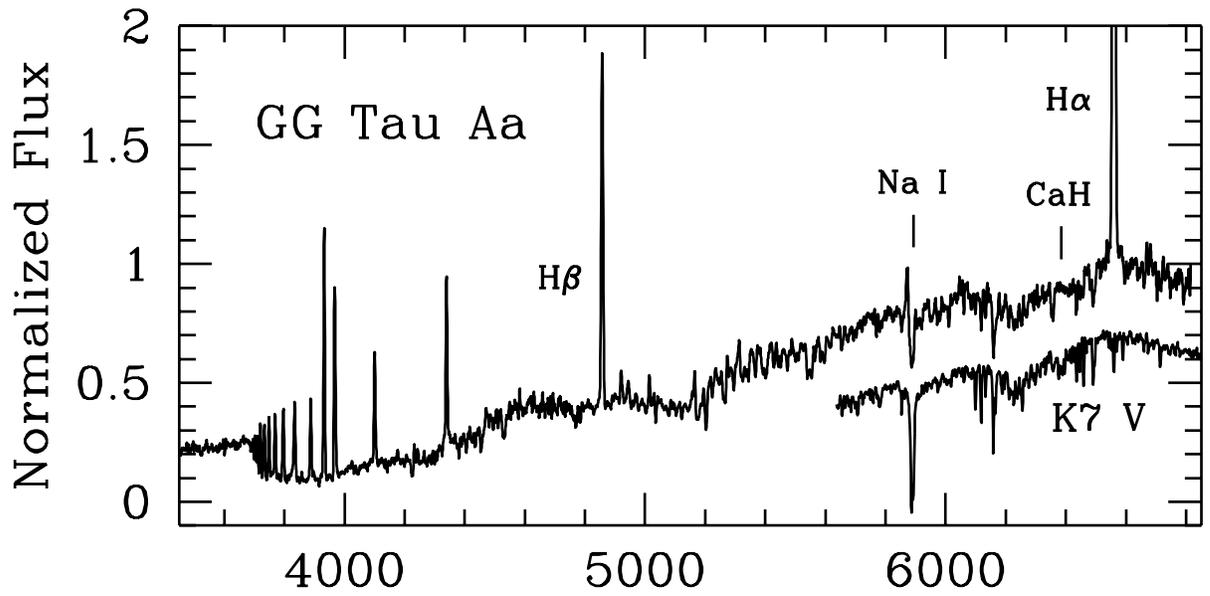
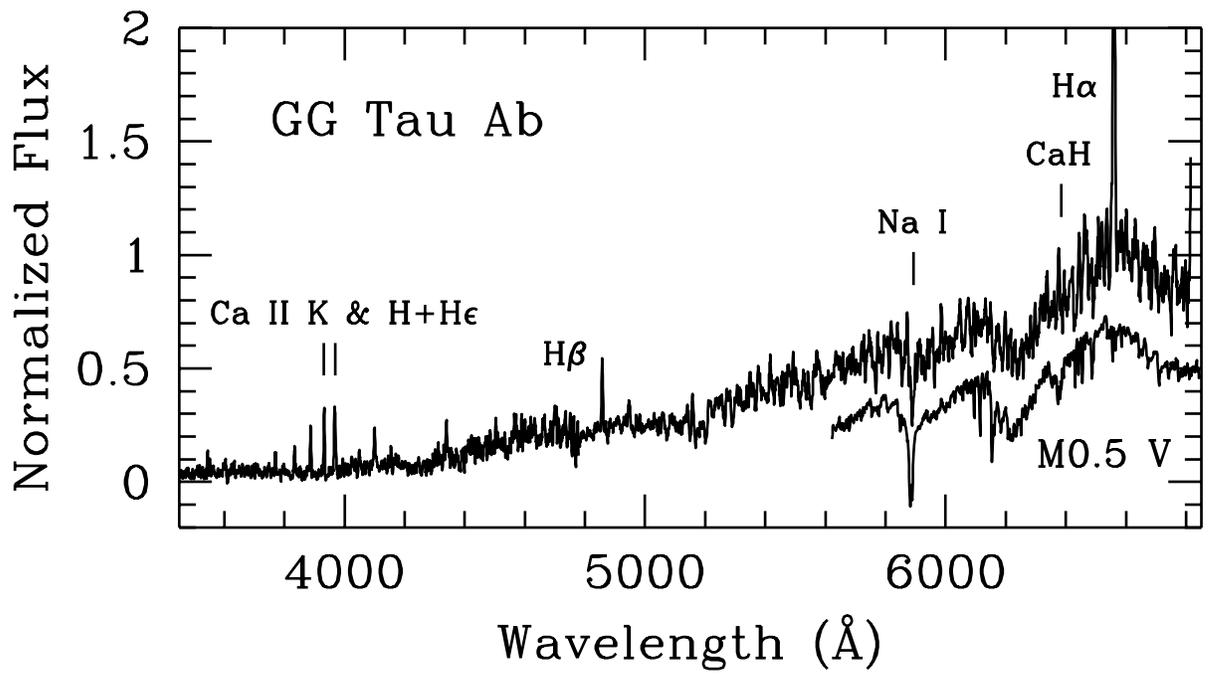
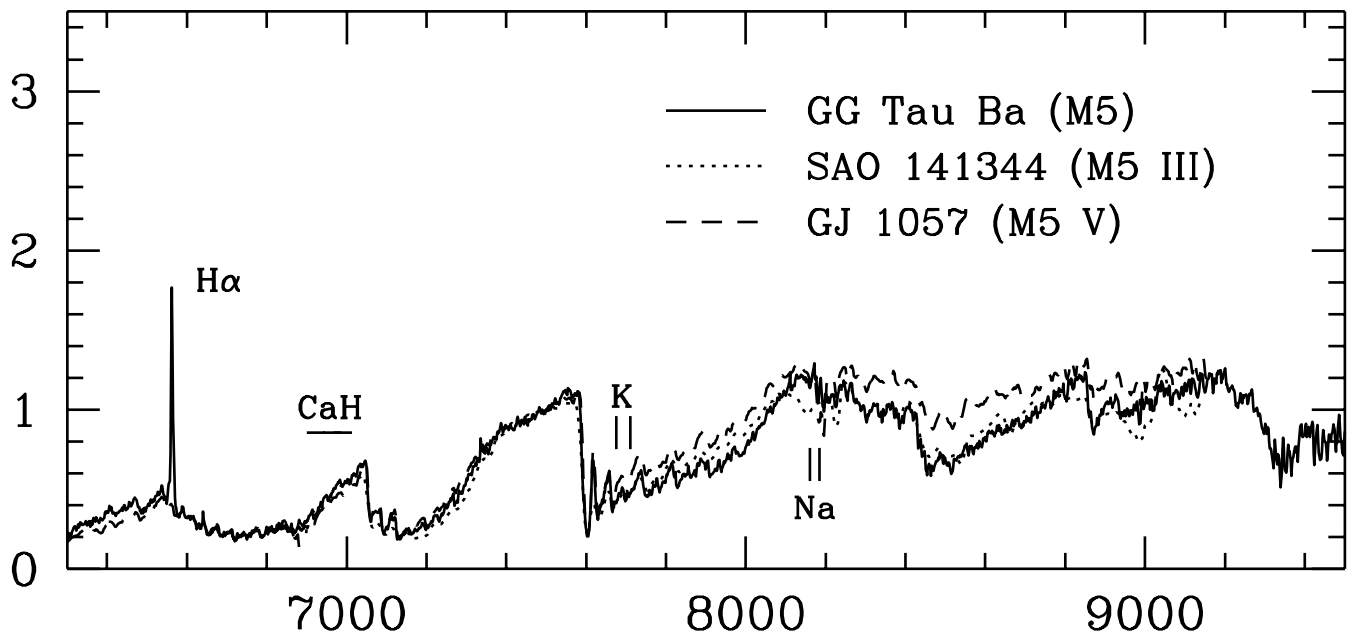
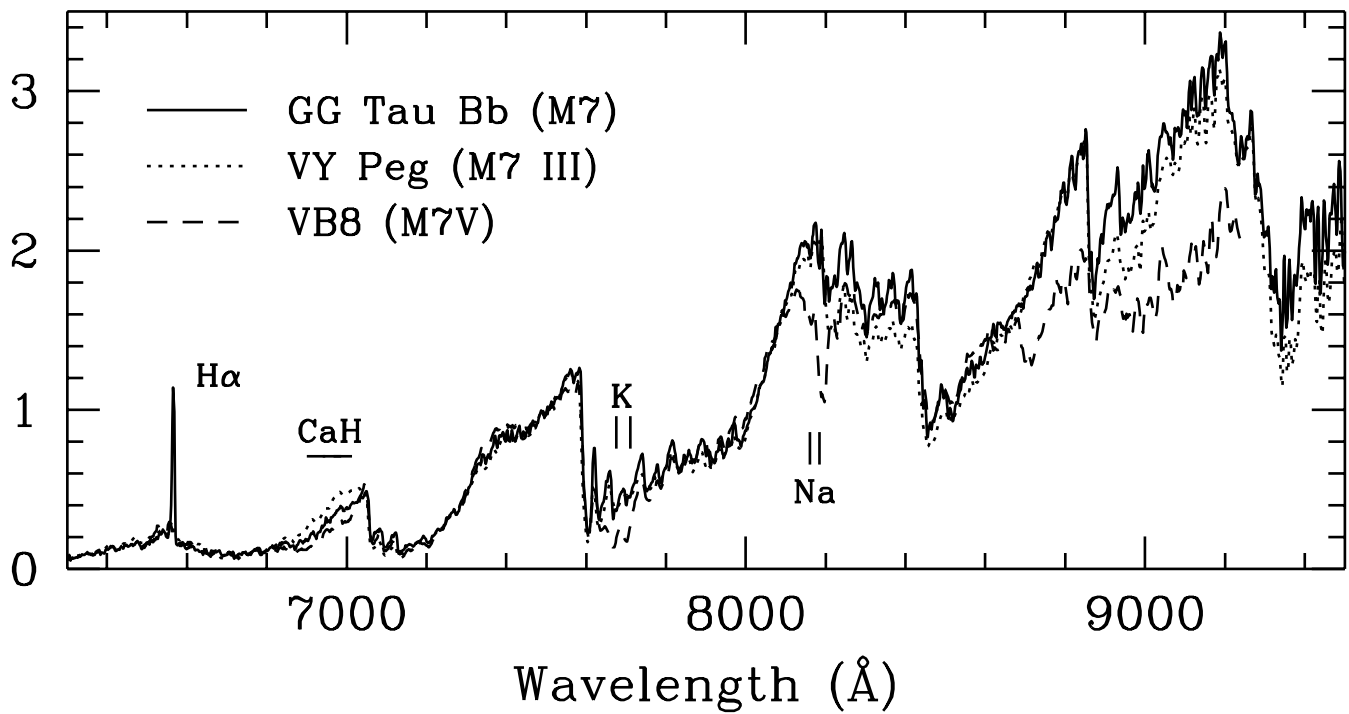
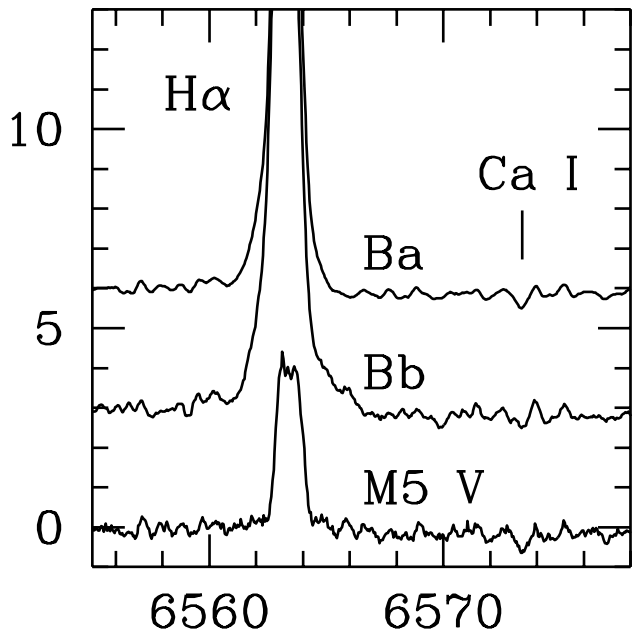
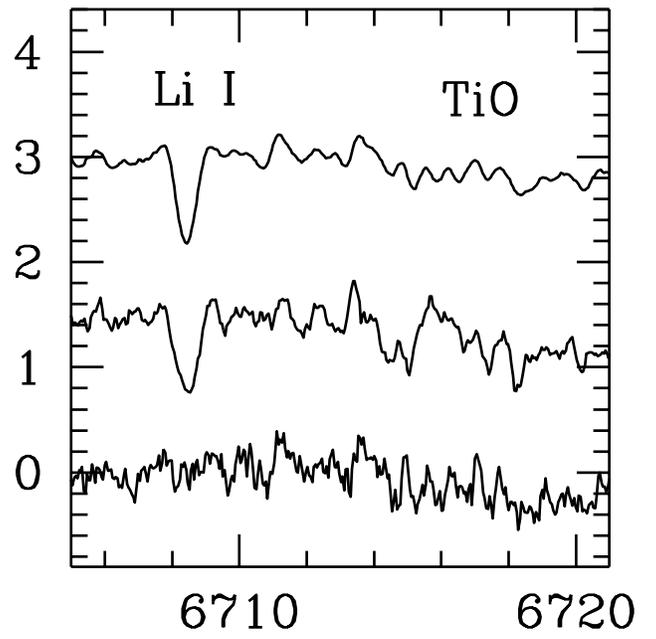
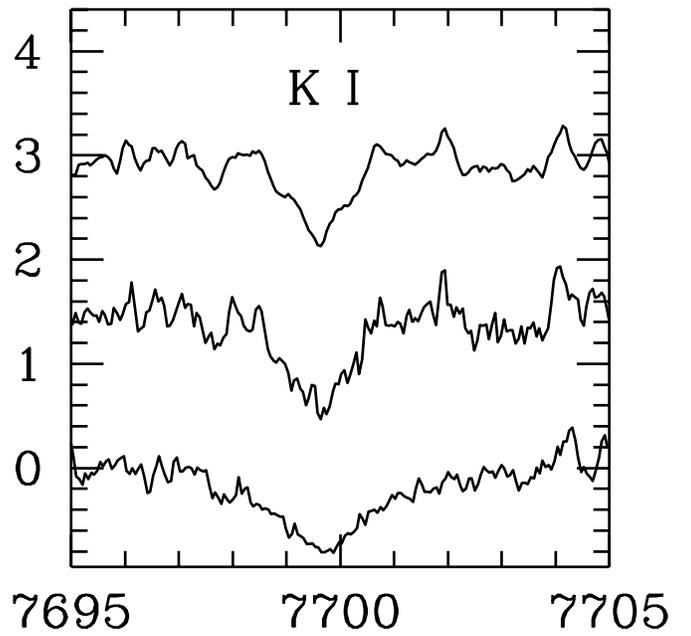
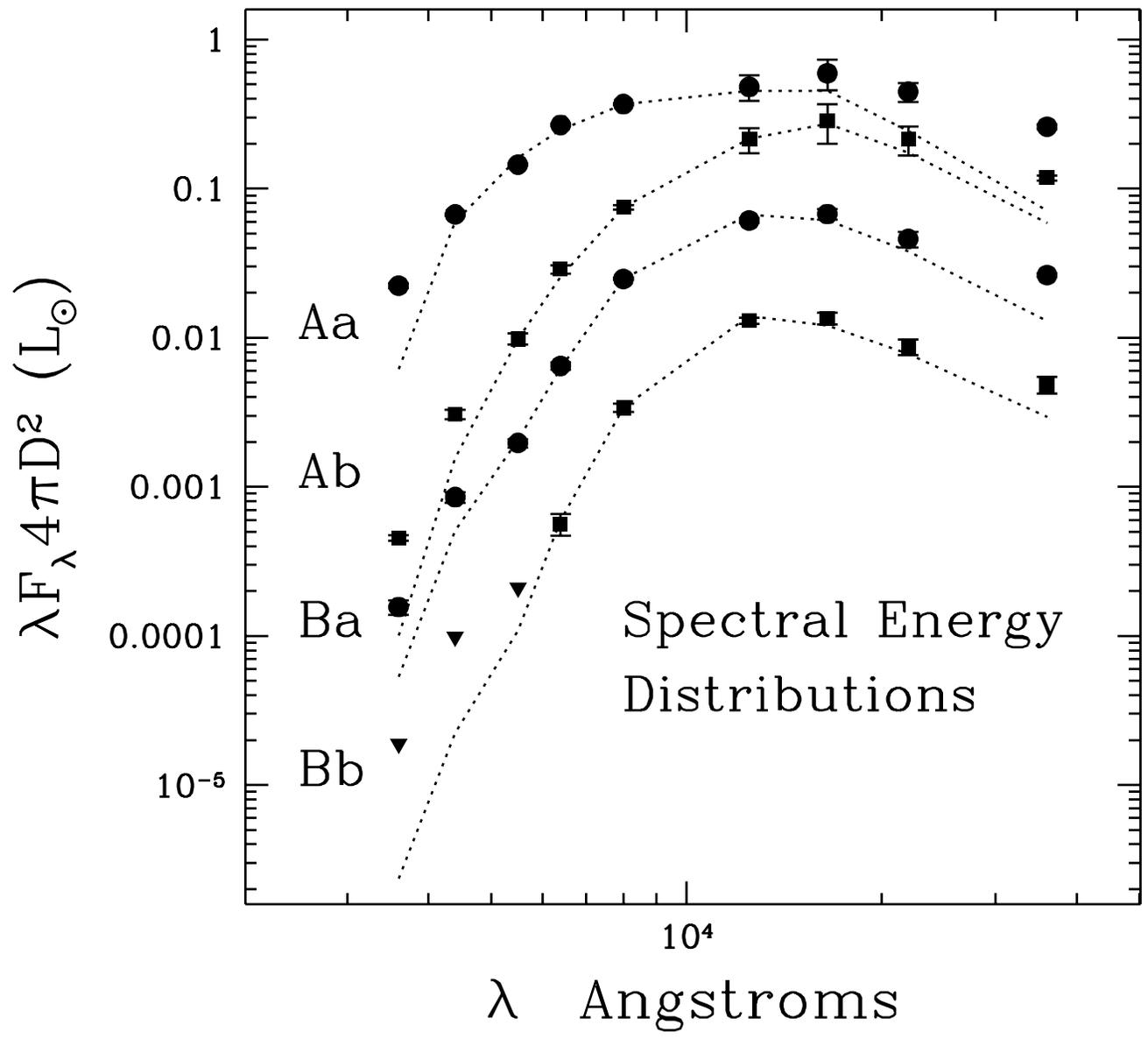
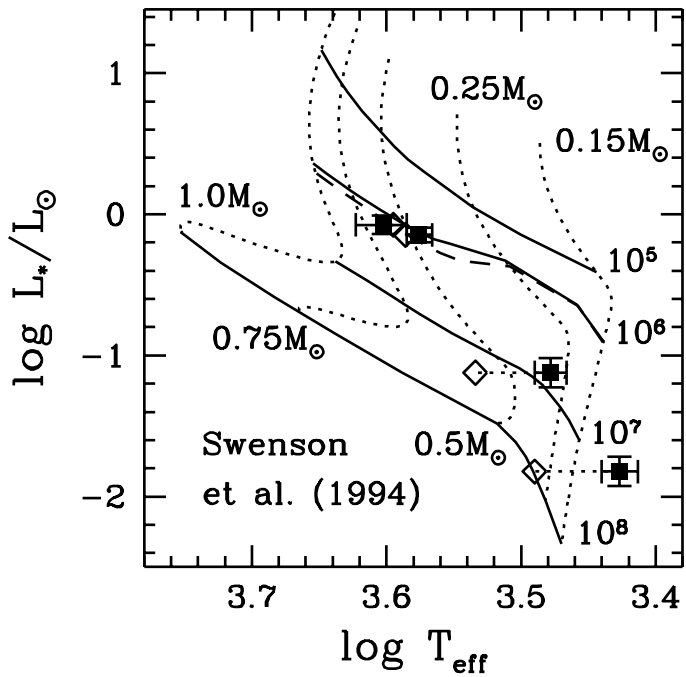
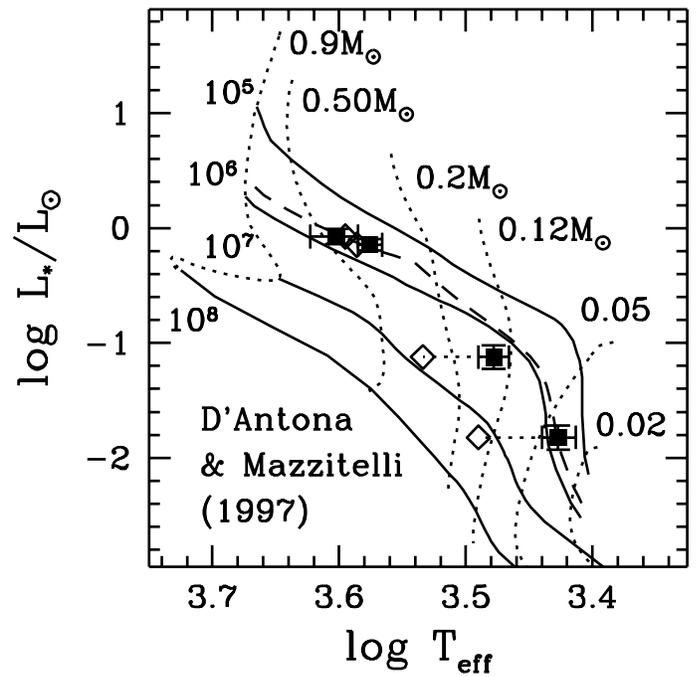
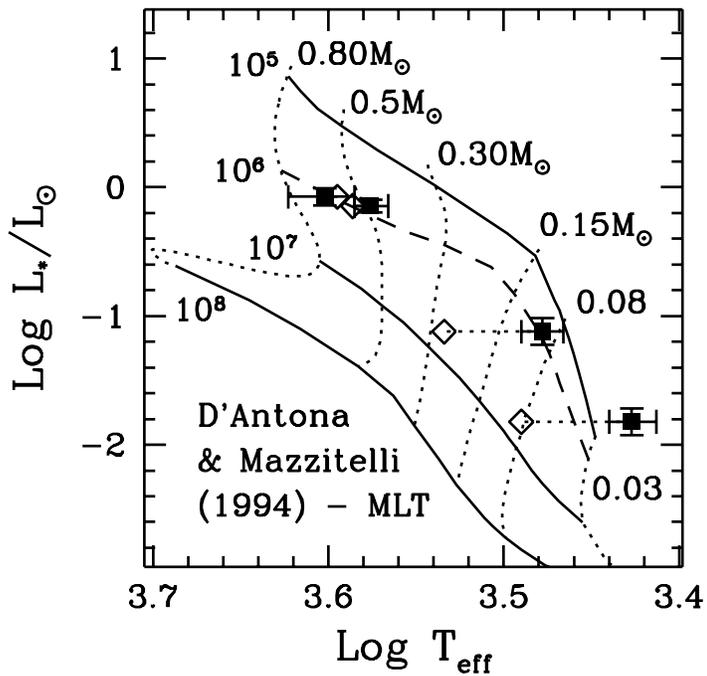
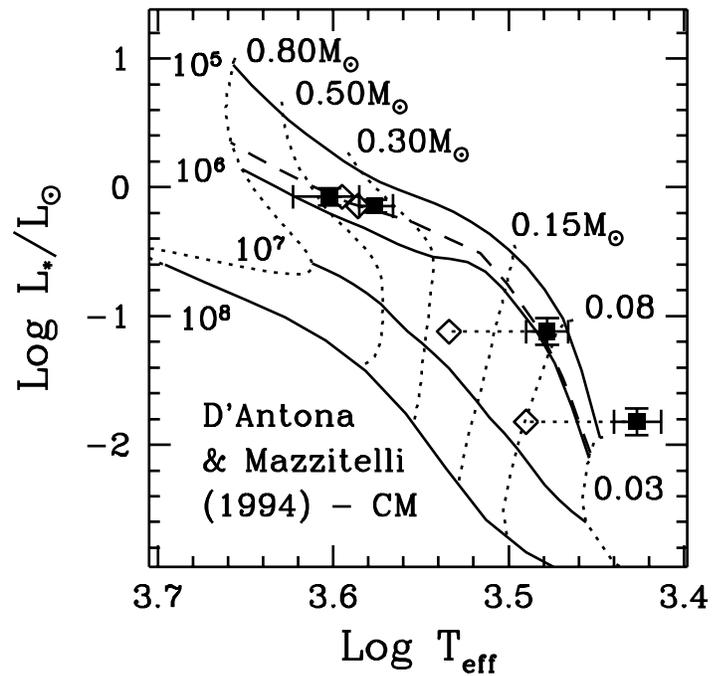
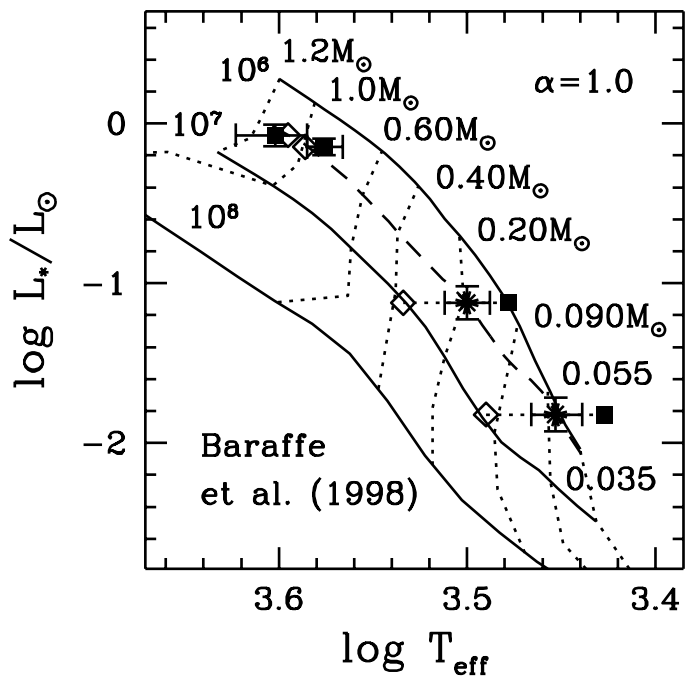
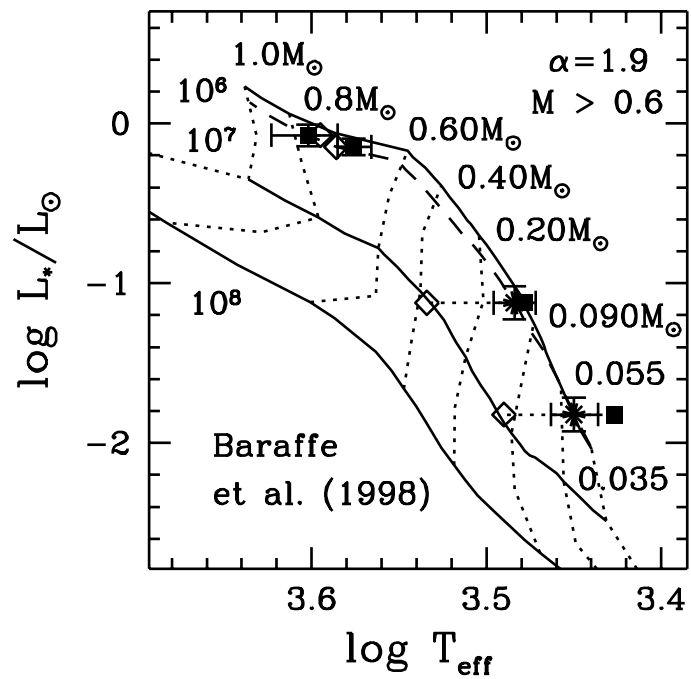
\end{document}